\documentclass[aps,pra,showpacs,superscriptaddress,preprint]{revtex4}
\usepackage{amsmath}
\usepackage{graphicx}

\catcode`ð=\active
 \defð{\u{g}}
 \catcode`Ð=\active
 \defÐ{\u{G}}
 \catcode`Ý=\active
\defÝ{\. I}
 \catcode`ö=\active
\defö{\"{o}}
 \catcode`Ö=\active
 \defÖ{\"O}
 \catcode`ü=\active
 \defü{\"{u}}
 \catcode`Ü=\active
 \defÜ{\"{U}}
 \catcode`Þ=\active
\defÞ{\c{S}}
 \catcode`þ=\active
 \defþ{\c{s}}
 \catcode`ý=\active
 \defý{{\i}}
 \catcode`ç=\active
\defç{\d{c}}
 \catcode`Ç=\active
\defÇ{\d{C}}

\input{tcilatex}

\begin{document}

\title{\textbf{An improved approximation scheme for the centrifugal term and
the Hul\textbf{t}h\'{e}n potential}}
\author{Sameer M. Ikhdair}
\email[E-mail: ]{sikhdair@neu.edu.tr}
\affiliation{Department of Physics, Near East University, Nicosia, North Cyprus, Turkey}
\date{%
\today%
}

\begin{abstract}
We present a new approximation scheme for the centrifugal term to solve the
Schr\"{o}dinger equation with the Hulth\'{e}n potential for any arbitrary $l$
state by means of a mathematical Nikiforov-Uvarov (NU) method. We obtain the
bound state energy eigenvalues and the normalized corresponding
eigenfunctions expressed in terms of the Jacobi polynomials or
hypergeometric functions for a particle exposed to this potential field. Our
numerical results of the energy eigenvalues are found to be in high
agreement with those results obtained by using the program based on a
numerical integration procedure. The $s$-wave ($l=0$) analytic solution for
the binding energies and eigenfunctions of a particle are also calculated.
The physical meaning of the approximate analytical solution is discussed.
The present approximation scheme is systematic and accurate.

Keywords: Bound states, Hulth\'{e}n potential, NU method, approximation
schemes.
\end{abstract}

\pacs{03.65.Ge, 12.39.Jh}
\maketitle

\newpage

\section{Introduction}

It is necessary to obtain the exact bound state energy spectrum of the Schr%
\"{o}dinger equations for some physical potential models. Therefore, much
works have been done to solve the wave equation for various radial and
angular potentials. Unfortunately, the exact analytic solutions (EAS) of
idealized quantum systems (QS), under consideration, are possible only in
the $s$-wave case with angular quantum number $l=0$ for some
exponential-type potential models. On the other hand, the Schr\"{o}dinger
equation cannot be solved analytically for $l\neq 0$ because of the
centrifugal term potential $\frac{l(l+1)}{r^{2}}.$ Over the past years, some
authors [1-16] have used the approximation $\frac{l(l+1)}{r^{2}}\approx 
\frac{l(l+1)\delta ^{2}e^{\delta r}}{\left( e^{\delta r}-1\right) ^{2}}$ for
the centrifugal term potential proposed by Greene and Aldrich [1] to obtain
the $l\neq 0$ analytic bound-states [2,4,5] and scattering states [7]
solutions of the non-relativistic [2,5] and relativistic [6] wave equations
with some exponential-type potentials such as Hulth\'{e}n potential [2-7],
Eckart potential [10-13], Manning-Rosen potential [14-16] and diatomic
molecular hyperbolical potential [17]. However, this approximation is valid
only for small values of the screening parameter $\delta $ and it breaks
down for large values of $\delta $ [5]. Therefore, there have been broad
interest and impressive efforts in order to find a new approximation scheme
which deals with the centrifugal term potential.

The Hulth\'{e}n potential [2,5,18] is the special case of the multiparameter
exponential-type potential model [19,20]. It takes the form%
\begin{equation}
V(r)=-\frac{V_{0}}{e^{\delta r}-1},\text{ }V_{0}=Ze^{2}\delta ,
\end{equation}%
where $V_{0}$ is a constant and $\delta $ is the screening parameter that
determines the range of the potential. If the potential is used for atoms,
then $V_{0}=Z\delta $ (in units $\hbar =c=e=1$), where $Z$ is identified as
the atomic number. The Hulth\'{e}n potential behaves like the Coulomb
potential near the origin $($i.e., $r\rightarrow 0)$ $V_{C}(r)=-Ze^{2}/r$ ,
but decreases exponentially in the asymptotic region when $r\gg 0,$ so its
capacity for bound states is smaller than the Coulomb potential [6,21-24].
This potential has been applied to a number of areas such as nuclear and
particle physics [25-27], atomic physics [28-31], molecular physics [32,33]
and chemical physics [34,35], etc.

The bound-state EAS of the Schr\"{o}dinger equation with the Hulth\'{e}n
potential can be solved in a closed form for $s$-waves (states with zero
orbital angular momentum $l$) [36]. However, for the case $l\neq 0,$ this
quantum system cannot be exactly solved. For implementing approximate
schemes economically and profitably; while dealing with practical quantum
mechanical problems, EAS of the Hulth\'{e}n potential is desirable although
nonperturbative and numerical solutions of different potentials may lead to
new physical ideas and/or calculational techniques in quantum physics. For
instance, the numerical integration of the Schr\"{o}dinger equation [37] is
used to obtain the energy eigenvalues numerically for the Hulth\'{e}n
potential case, this provides a probe and/or test for the exactness of any
analytic solution. One-parameter variational calculations are carried out in
such numerical integration methods. The variational results are practically
identical to the exact energies, except in the high-screening region. These
variational calculations turn to become sophisticated in the solution of Schr%
\"{o}dinger equation with multi-parameter potentials. However, no
\textquotedblright exact\textquotedblright\ values obtained from a numerical
integration of the Schr\"{o}dinger equation have been available to assess
the accuracy of the various methods [37]. Hence, it is important to note
that the analytic solution of any quantum potential model, even if it is an
approximated solution, is indispensable since the obtained expressions for
energy eigenvalues and eigen functions contain all the necessary information
regarding the quantum system under consideration.

In the non-relativistic case, for $l\neq 0,$ several techniques have been
used to obtain approximate analytic solutions, some authors have obtained
the bound-state energy eigenvalues by using the numerical integration
approach [37,38], quasi-analytical variational [37,39], perturbation [40],
SUSYQM [3], shifted $1/N$ expansion [41], AIM [5] and Nikiforov-Uvarov (NU)
[2] methods. The results obtained by some of these methods [3,5] are in good
agreement with the numerical integration approach [37] for low-screening
region (small values of the screening parameter $\delta $) but the agreement
becomes poor in the high-screening regime [5].

Recently, Haouat and Chetouani [42] have solved the Klein-Gordon and Dirac
equations in the presence of the Hulth\'{e}n potential, where the energy
spectrum and the scattering wavefunctions are obtained for spin-$0$ and spin-%
$\frac{1}{2}$ particles, making a slight modification to the usual
approximation scheme, $\frac{1}{r^{2}}\approx \alpha ^{2}\frac{e^{-\gamma
\alpha r}}{\left( 1-e^{-\alpha r}\right) ^{2}}$ where $\gamma $ is a
dimensionless parameter ($\gamma =0,1$ and $2$) for the centrifugal term
potential. They found that the good approximation, however, when the
screening parameter $\alpha $ and the dimensionless parameter $\gamma $ are
taken as $\alpha =0.1$ and $\gamma =$1, respectively, which is simply the
case of the normal approximation used in the literature. Also, Jia and
collaborators [43] have recently proposed a new alternative approximation
scheme, $\frac{1}{r^{2}}\approx \alpha ^{2}\left( \frac{\omega }{e^{\alpha
r}-1}+\frac{1}{\left( e^{\alpha r}-1\right) ^{2}}\right) $ where $\omega $
is a dimensionless parameter ($\omega =1.030$)$,$ for the centrifugal
potential to improve the numerical energy eigenvalues of the Hulth\'{e}n
potential. When taking $\omega =1,$ their approximation can be reduced to
the usual approximation [1-16]. However, the accuracy of their numerical
results [43] is still in poor agreement with the other numerical integration
and variational methods [37] especially in high-screening $\delta $ regime.
This problem could be solved by making a better approximation for the
centrifugal term potential. In this work, for any arbitrary $l$-state, we
aim to obtain approximate energy eigenvalues and corresponding normalized
wave functions for the Hulth\'{e}n potential in high agreement with the
numerical method [37]. Hence, we present an alternative effective
approximation that gives highly accurate numerical energy eigenvalues of the
Hulth\'{e}n potential as a function of screening parameter for all states
with $Z=1.$

This paper is organized as follows: In the next Section, the NU method is
briefly introduced. In Section 3, the $l$-states Schr\"{o}dinger equation
for the Hulth\'{e}n potential is solved within the new effective
approximation scheme and using the NU method. The calculated energy
eigenvalues and wave functions are compared with the other ones found by
using different analytical and numerical methods. The normalized wave
functions are obtained in Section 4. Finally, the relevant conclusions are
given in Section 5.

\section{NU Method}

The NU method is briefly introduced here and the details can be found in
Nikiforov-Uvarov handbook [44]. This method was proposed to solve the
second-order differential wave equation of the hypergeometric-type:%
\begin{equation}
\psi _{n}^{\prime \prime }(s)+\frac{\widetilde{\tau }(s)}{\sigma (s)}\psi
_{n}^{\prime }(s)+\frac{\widetilde{\sigma }(s)}{\sigma ^{2}(s)}\psi
_{n}(s)=0,
\end{equation}%
where $\sigma (s)$ and $\widetilde{\sigma }(s)$ are polynomials, at most of
second-degree, and $\widetilde{\tau }(s)$ is a first-degree polynomial. The
prime sign denotes the differentiation with respect to $s.$ To find a
particular solution of Eq. (2), one can decompose the wavefunction $\psi
_{n}(z)$ as follows: 
\begin{equation}
\psi _{n}(s)=\phi _{n}(s)y_{n}(s),
\end{equation}%
leading to a hypergeometric type equation%
\begin{equation}
\sigma (s)y_{n}^{\prime \prime }(s)+\tau (s)y_{n}^{\prime }(s)+\lambda
y_{n}(s)=0.
\end{equation}%
The first part of the wavefunctions $\psi _{n}(s)$ is the solution of the
differential equation,%
\begin{equation}
\sigma (s)\phi ^{\prime }(s)-\pi (s)\phi (s)=0,
\end{equation}%
where%
\begin{equation}
\tau (s)=\widetilde{\tau }(s)+2\pi (s),
\end{equation}%
and $\lambda $ in (4) is a parameter defined as%
\begin{equation}
\lambda =\lambda _{n}=-n\tau ^{\prime }(s)-\frac{n\left( n-1\right) }{2}%
\sigma ^{\prime \prime }(s),\text{ }n=0,1,2,\cdots .
\end{equation}%
The $\tau (s)$ is a polynomial function of the parameter $s$ whose first
derivative $\tau ^{\prime }(s)$ must be negative which is the essential
condition in choosing the proper solutions. The second part of the
wavefunctions (3) is a hypergeometric-type function obtained by Rodrigues
relation:%
\begin{equation}
y_{n}(s)=\frac{B_{n}}{\rho (s)}\frac{d^{n}}{ds^{n}}\left[ \sigma ^{n}(s)\rho
(s)\right] ,
\end{equation}%
where $B_{n}$ is a constant related to normalization and the weight function 
$\rho (s)$ can be found by [44]%
\begin{equation}
\sigma (s)\rho ^{\prime }(s)+\left[ \sigma ^{\prime }(s)-\tau (s)\right]
\rho (s)=0,
\end{equation}%
The function $\pi (s)$ and the parameter $\lambda $ are defined as%
\begin{equation}
\pi (s)=\frac{\sigma ^{\prime }(s)-\widetilde{\tau }(s)}{2}\pm \sqrt{\left( 
\frac{\sigma ^{\prime }(s)-\widetilde{\tau }(s)}{2}\right) ^{2}-\widetilde{%
\sigma }(s)+k\sigma (s)},
\end{equation}%
\begin{equation}
\lambda =k+\pi ^{\prime }(s),
\end{equation}%
where $\pi (s)$ has to be a polynomial of degree at most one. The
discriminant under the square root sign in Eq. (10) must be set to zero and
then has to be solved for $k$ [44]. Finally, the energy eigenvalue equation
is simply found by solving Eqs. (7) and (11).

\section{Bound State Solutions}

The Schr\"{o}dinger equation for a central molecular potential $V(r)$ can be
written as 
\begin{equation}
\left( \frac{\hbar ^{2}}{2\mu }\nabla ^{2}+E_{nl}-V(r)\right) \psi _{nlm}(%
\mathbf{r,}\theta ,\phi )=0,\text{ }
\end{equation}%
where the representation of the Laplacian operator $\nabla ^{2},$ in
spherical coordinates, is 
\begin{equation}
\nabla ^{2}=\frac{\partial ^{2}}{\partial r^{2}}+\frac{2}{r}\frac{\partial }{%
\partial r}+\frac{1}{r^{2}}\left( \frac{1}{\sin \theta }\frac{\partial }{%
\partial \theta }\sin \theta \frac{\partial }{\partial \theta }+\frac{1}{%
\sin ^{2}\theta }\frac{\partial ^{2}}{\partial \phi ^{2}}\right) .
\end{equation}%
Here the wave functions $\psi _{nlm}(\mathbf{r,}\theta ,\phi )$ belong to
the energy eigenvalues $E_{nl}$ and $V(r)$ stands for the molecular
potential in the configuration space and $r$ represents the
three-dimensional intermolecular distance $\left(
\sum_{i=1}^{3}x_{i}^{2}\right) ^{1/2}.$ Let us decompose the wave function
in (12) as follows:%
\begin{equation}
\psi _{nlm}(\mathbf{r,}\theta ,\phi )=r^{-1}u_{nl}(r)Y_{lm}(\theta ,\phi ),
\end{equation}%
where $Y_{lm}(\theta ,\phi )$ represents contribution from the
hyperspherical harmonics that arise in higher dimensions. Substituting the
wave functions (14) into Eq. (12), the result is [45,46]%
\begin{equation}
\left\{ \frac{d^{2}}{dr^{2}}-\frac{l\left( l+1\right) }{r^{2}}+\frac{2\mu }{%
\hbar ^{2}}\left[ E_{nl}-V(r)\right] \right\} u_{n,l}(r)=0,
\end{equation}%
where $E_{nl}$ is the bound-state energy of the system under consideration,
i.e., $E_{nl}<0$ and the term $\frac{l\left( l+1\right) }{r^{2}}$ is known
as the centrifugal term$.$ We also should be careful about the behavior of
the wave function $u_{nl}(r)$ near $r=0$ and $r\rightarrow \infty .$
Furthermore, $u_{nl}(r)$ should be normalizable [47].

We can rewrite Eq. (15) for the Hulth\'{e}n potential as%
\begin{equation}
\frac{d^{2}u_{nl}(r)}{dr^{2}}+\left[ \frac{2\mu E_{nl}}{\hbar ^{2}}+\frac{%
2\mu Ze^{2}\delta }{\hbar ^{2}}\frac{e^{-\delta r}}{1-e^{-\delta r}}-\frac{%
l(l+1)}{r^{2}}\right] u_{nl}(r)=0,
\end{equation}%
where $E_{nl}$ is the bound state energy of the system and $n$ and $l$
signify the radial and angular quantum numbers, respectively. When $l=0$ ($s$%
-wave), Eq. (16) with the Hulth\'{e}n potential can be exactly solved
[36,48-50], but for the case $l\neq 0,$ Eq. (16) cannot be exactly solved.
So we must find a new approximation to the entrifugal term to solve the
equation analytically. The new proposed approximation is based on the
expansion of the centrifugal term in a series of exponentials depending on
the intermolecular distance $r$ and keeping terms up to second order. For
small $0.4\leq \delta r\leq 1.2$ [5] (i.e., small screening parameter $%
\delta ),$ Eq. (16) is very well approximated to centrifugal term. However,
for large screening parameter, a better approximation to the centrifugal
term should be made. Hence, instead of employing the usual approximation
given in [1-16], we propose an alternative approximation scheme casted in
the form: 
\begin{equation*}
\frac{1}{r^{2}}\approx \delta ^{2}\left[ d_{0}+v(r)+v^{2}(r)\right] ,\text{ }%
v(r)=\frac{e^{-\delta r}}{1-e^{-\delta r}},
\end{equation*}%
\begin{equation}
\frac{1}{r^{2}}\approx \delta ^{2}\left[ d_{0}+\frac{1}{e^{\delta r}-1}+%
\frac{1}{\left( e^{\delta r}-1\right) ^{2}}\right] ,
\end{equation}%
for the centrifugal term which takes a similar ans\"{a}tze like the Hulth%
\'{e}n potential. Under the coordinate transformation $r\rightarrow x,$ it
is convenient to shift the origin by defining $x=(r-r_{0})/r_{0},$ we obtain%
\begin{equation}
\left( 1+x\right) ^{-2}=\gamma ^{2}\left[ d_{0}+\frac{1}{e^{\gamma (1+x)}-1}+%
\frac{1}{\left( e^{\gamma (1+x)}-1\right) ^{2}}\right] ,\text{ }\gamma
=r_{0}\delta .
\end{equation}%
Further, expanding Eq. (17) around $r=r_{0}$ $(x=0),$ we obtain the
following expansion: 
\begin{equation*}
1-2x+O(x^{2})=\gamma ^{2}\left( d_{0}+\frac{1}{e^{\gamma }-1}+\frac{1}{%
\left( e^{\gamma }-1\right) ^{2}}\right)
\end{equation*}%
\begin{equation}
-\gamma ^{3}\left( \frac{1}{e^{\gamma }-1}+\frac{3}{\left( e^{\gamma
}-1\right) ^{2}}+\frac{2}{\left( e^{\gamma }-1\right) ^{3}}\right)
x+O(x^{2}),
\end{equation}%
from which we have

\begin{equation*}
\gamma ^{2}\left[ d_{0}+\frac{1}{e^{\gamma }-1}+\frac{1}{(e^{\gamma }-1)^{2}}%
\right] =1,
\end{equation*}%
\begin{equation}
\gamma ^{3}\left( \frac{1}{e^{\gamma }-1}+\frac{3}{\left( e^{\gamma
}-1\right) ^{2}}+\frac{2}{\left( e^{\gamma }-1\right) ^{3}}\right) =2.
\end{equation}%
Therefore, the shifting parameter $d_{0}$ is to be found from the solution
of the above two equations as 
\begin{equation}
d_{0}=\frac{1}{\gamma ^{2}}-\frac{1}{e^{\gamma }-1}-\frac{1}{(e^{\gamma
}-1)^{2}}=0.0823058167837972,
\end{equation}%
where $e$ is the base of the natural logarithms, $e=2.718281828459045$ and
the parameter $\gamma =0.4990429999.$

Therefore, we may cast the centrifugal term as%
\begin{equation}
\underset{\delta \rightarrow 0}{\lim }\delta ^{2}\left[ \frac{1}{\gamma ^{2}}%
-\frac{1}{e^{\gamma }-1}-\frac{1}{(e^{\gamma }-1)^{2}}+\frac{e^{-\delta r}}{%
1-e^{-\delta r}}+\left( \frac{e^{-\delta r}}{1-e^{-\delta r}}\right) ^{2}%
\right] =\frac{1}{r^{2}}.
\end{equation}%
To conclude, it is important to note that when $d_{0}=0,$ the approximation
expression (17) is reduced to the usual approximation used in literature
[1-16]. The variation of the centrifugal term potential $l(l+1)/r^{2}$ and
the proposed approximation expression given in (17) versus $\delta r$ are
plotted in Figure 1. Obviously, the approximate centrifugal term potential
(17) and $l(l+1)/r^{2}$ are similar and coincide in both high-screening as
well as in the low-screening regimes as shown in Figure 1.

Inserting the approximation expression (17) into Eq. (16) and changing the
variables $r\rightarrow s=e^{-\delta r}$ through the mapping function $%
s=f(r),$ where $r\in \lbrack 0,\infty )$ or $s\in \lbrack 1,0],$ leads us to
obtain the following equation

\begin{equation}
u_{nl}{}^{\prime \prime }(s)+\frac{(1-s)}{s(1-s)}u_{nl}^{\prime }(s)+\frac{1%
}{\left[ s(1-s)\right] ^{2}}\left[ -\varepsilon
_{nl}^{2}+(c_{1}-c_{2}+2\varepsilon _{nl}^{2})s-(c_{1}+\varepsilon
_{nl}^{2})s^{2}\right] u_{nl}(s)=0,
\end{equation}%
where%
\begin{equation}
\varepsilon _{nl}=\sqrt{\Delta E_{l}-\frac{2\mu E_{nl}}{\hbar ^{2}\delta ^{2}%
}},\text{ \ }\Delta E_{l}=l(l+1)d_{0},\text{ }c_{1}=\frac{2\mu Ze^{2}}{\hbar
^{2}\delta },\text{ }c_{2}=l(l+1).
\end{equation}%
In the present work, we will deal with bound state solutions, i.e., the
radial part of the wavefunction $\psi _{nlm}(\mathbf{r,}\theta ,\phi )$ must
satisfy the boundary condition that $u_{nl}(r)/r$ becomes zero when $%
r\rightarrow \infty ,$ and $u_{nl}(r)/r$ is finite at $r=0.$ In addition, we
require $E_{nl}\leq \frac{\hbar ^{2}\delta ^{2}}{2\mu }\Delta E_{l},$ i.e., $%
\varepsilon _{nl}\geq 0$ [36,51-53]. Comparing Eqs. (23) and (2), we obtain
the relevant polynomials:%
\begin{equation}
\widetilde{\tau }(s)=1-s,\sigma (s)=s(1-s),\widetilde{\sigma }%
(r)=-\varepsilon _{nl}^{2}+(c_{1}-c_{2}+2\varepsilon
_{nl}^{2})s-(c_{1}+\varepsilon _{nl}^{2})s^{2}.
\end{equation}%
Inserting the polynomials given by Eq. (25) into Eq. (10) gives the
polynomial:%
\begin{equation}
\pi (s)=-\frac{s}{2}\pm \frac{1}{2}\sqrt{\widetilde{a}s^{2}+\widetilde{b}s+%
\widetilde{c}},
\end{equation}%
where $\widetilde{a}=1+4(c_{1}+\varepsilon _{nl}^{2}-k),$ $\widetilde{b}%
=4(k-c_{1}+c_{2}-2\varepsilon _{nl}^{2})$ and $\widetilde{c}=4\varepsilon
_{nl}^{2}.$ The equation of quadratic form under the square root sign of Eq.
(26) must be solved by setting the discriminant of this quadratic equal to
zero: $\Delta =\widetilde{b}^{2}-4\widetilde{a}\widetilde{c}=0.$ This
discriminant gives a new quadratic equation can be solved for the constant $%
k $ to obtain the two roots:%
\begin{equation}
k_{1,2}=c_{1}-c_{2}\pm \varepsilon _{nl}\sqrt{1+4c_{2}}.
\end{equation}%
When the two values of $k$ given in Eq. (27) are substituted into Eq. (26),
the four possible forms of $\pi (s)$ are obtained as%
\begin{equation}
\pi (s)=-\frac{s}{2}\pm \left\{ 
\begin{array}{cc}
\left[ \left( \varepsilon _{nl}-\frac{1}{2}\sqrt{1+4c_{2}}\right)
s-\varepsilon _{nl}\right] & \text{for }k_{1}=c_{1}-c_{2}+\varepsilon _{nl}%
\sqrt{1+4c_{2}}, \\ 
\left[ \left( \varepsilon _{nl}+\frac{1}{2}\sqrt{1+4c_{2}}\right)
s-\varepsilon _{nl}\right] & \text{\ \ for }k_{2}=c_{1}-c_{2}-\varepsilon
_{nl}\sqrt{1+4c_{2}}.%
\end{array}%
\right.
\end{equation}%
One of the four values of the polynomial $\pi (s)$ is just proper to obtain
the bound state energy states because $\tau (s)$ given by Eq. (6) has a
negative derivative for this value of $\pi (s)$ [44]. Therefore, the most
suitable expression of $\pi (s)$ is chosen as%
\begin{equation}
\pi (s)=-\frac{s}{2}-\left[ \left( \varepsilon _{nl}+\frac{1}{2}\sqrt{%
1+4c_{2}}\right) s-\varepsilon _{nl}\right] ,
\end{equation}%
for $k_{2}=c_{1}-c_{2}-\varepsilon _{nl}\sqrt{1+4c_{2}}.$ Hence, $\tau (s)$
and $\tau ^{\prime }(s)$ are obtained%
\begin{equation*}
\tau (s)=1+2\varepsilon _{nl}-2\left[ 1+\varepsilon _{nl}+\frac{1}{2}\sqrt{%
1+4c_{2}}\right] s,
\end{equation*}%
\begin{equation}
\tau ^{\prime }(s)=-2\left[ 1+\varepsilon _{nl}+\frac{1}{2}\sqrt{1+4c_{2}}%
\right] ,
\end{equation}%
where $\tau ^{\prime }(s)$ represents the derivative of $\tau (s).$ Using
Eqs. (25), (29) and (30), the following expressions for $\lambda $ and $%
\lambda _{n}$ are obtained, respectively,%
\begin{equation}
\lambda =\lambda _{n}=n^{2}+\left[ 1+2\varepsilon _{nl}+\sqrt{1+4c_{2}}%
\right] n,\text{ }(n=0,1,2,\cdots ),
\end{equation}%
\begin{equation}
\lambda =c_{1}-c_{2}-\frac{1}{2}(1+2\varepsilon _{nl})\left[ 1+\sqrt{1+4c_{2}%
}\right] ,
\end{equation}%
where $n$ is the number of nodes of the radial wave function $u_{n,l}(r)$.
When $\lambda =\lambda _{n},$ an expression for $\varepsilon _{nl}$ is
obtained as%
\begin{equation}
\varepsilon _{nl}=\frac{c_{1}}{2(n+l+1)}-\frac{\left( n+l+1\right) }{2},%
\text{ }(n,l=0,1,2,\cdots ).
\end{equation}%
Also, with the aid of Eq. (24), the previous energy equation gives the
following bound state energy eigenvalue equation:%
\begin{equation}
E_{nl}=\frac{\hbar ^{2}\delta ^{2}}{2\mu }\left\{ l(l+1)d_{0}-\left[ \frac{%
\mu Ze^{2}}{\hbar ^{2}\delta \left( n+l+1\right) }-\frac{\left( n+l+1\right) 
}{2}\right] ^{2}\right\} .
\end{equation}%
In the case of the $s$-wave ($l=0$), the previous equation turns to be

\begin{equation}
E_{n}=-\frac{\hbar ^{2}\delta ^{2}}{2\mu }\left( \frac{\mu Ze^{2}}{\hbar
^{2}\delta \left( n+1\right) }-\frac{n+1}{2}\right) ^{2},
\end{equation}%
which is identical to the ones obtained before using the factorization
method [36], SUSYQM approach [3,28,54], quasi-linearization method [55] and
NU method [2,6,32]. Further, if we take the shift parameter $d_{0}=0$ in the
present approximation, Eq. (34) reduces to

\begin{equation}
E_{nl}=-\frac{\hbar ^{2}\delta ^{2}}{2\mu }\left[ \frac{\mu Ze^{2}}{\hbar
^{2}\delta \left( n+l+1\right) }-\frac{\left( n+l+1\right) }{2}\right] ^{2},
\end{equation}%
which is also identical with the energy eigenvalues formula given in Eq.
(32) of Ref. [5], Eq. (24) of Ref. [43] and Eq. (28) of Ref. [2].

Let us turn to the calculations of the wave function $y_{n}(s),$ which is
the first part solution of hypergeometric-type equation, we need to multiply
Eq. (4) by the weight function $\rho (s)$ so that it can be rewritten in
self-adjoint form [44]%
\begin{equation}
\left[ \omega (s)y_{n}^{\prime }(s)\right] ^{\prime }+\lambda \rho
(s)y_{n}(s)=0.
\end{equation}%
The weight function $\rho (s)$ that satisfies Eqs. (9) and (37) is found as%
\begin{equation}
\rho (s)=s^{2\varepsilon _{nl}}(1-s)^{(2l+1)},
\end{equation}%
which gives the Rodrigues relation (8):%
\begin{equation}
y_{nl}(s)=B_{nl}s^{-2\varepsilon _{nl}}(1-s)^{-(2l+1)}\frac{d^{n}}{ds^{n}}%
\left[ s^{n+2\varepsilon _{nl}}(1-s)^{n+2l+1}\right] =B_{nl}P_{n}^{(2%
\varepsilon _{nl},2l+1)}(1-2s).
\end{equation}%
Further, inserting the values of $\sigma (s)$ and $\pi (s)$ given in Eqs.
(25) and (29) into Eq. (5), one can find the other part of the wave function
as%
\begin{equation}
\phi (s)=s^{\varepsilon _{nl}}(1-s)^{(l+1)}.
\end{equation}%
Hence, the wave functions in Eq. (3) become%
\begin{equation}
u_{nl}(s)=\mathcal{N}_{nl}s^{\varepsilon
_{nl}}(1-s)^{l+1}P_{n}^{(2\varepsilon _{nl},2l+1)}(1-2s),\text{ }s\in
\lbrack 1,0)
\end{equation}%
where $\mathcal{N}_{nl}$ is the normalization constant to be determined in
the next section. Finally, the unnormalized radial wave functions are
obtained as%
\begin{equation}
\psi _{nlm}(\mathbf{r,}\theta ,\phi )=\mathcal{N}_{nl}r^{-1}\left(
e^{-\delta r\text{ }}\right) ^{\varepsilon _{nl}}(1-e^{-\delta
r})_{2}^{l+1}F_{1}(-n,n+2\left( \varepsilon _{nl}+l+1\right) ;2\varepsilon
_{nl}+1;e^{-\delta r})Y_{lm}(\theta ,\phi ).
\end{equation}%
Thus, the Jacobi polynomials can be expressed in terms of the hypergeometric
functions [56]:%
\begin{equation}
P_{n}^{\left( a,b\right) }(1-2x)=_{2}F_{1}(-n,n+a+b+1;a+1;x),
\end{equation}
where $_{2}F_{1}(a,b;c;x)=\frac{\Gamma (c)}{\Gamma (a)\Gamma (b)}%
\dsum\limits_{k=0}^{\infty }$ $\frac{\Gamma (a+k)\Gamma (b+k)}{\Gamma (c+k)}%
\frac{x^{k}}{k!}.$ The hypergeometric function $_{2}F_{1}(a,b;c;x)$ is a
special case of the generalized hypergeometric function [56]%
\begin{equation}
_{p}F_{q}(\alpha _{1},\alpha _{2},\cdots ,\alpha _{p};\beta _{1},\beta
_{1},\cdots ,\beta _{q};x)=\dsum\limits_{k=0}^{\infty }\frac{\left( \alpha
_{1}\right) _{k}\left( \alpha _{2}\right) _{k}\cdots \left( \alpha
_{p}\right) }{\left( \beta _{1}\right) _{k}\left( \beta _{2}\right)
_{k}\cdots \left( \beta _{q}\right) }\frac{x^{k}}{k!},
\end{equation}
where the Pochhammer symbol is defined by $\left( y\right) _{k}=\Gamma
(y+k)/\Gamma (y).$

In the case $l=0,$ the above wave functions become

\begin{equation}
\psi _{n}(\mathbf{r)}=D_{n}r^{-1}\left( e^{-\delta r\text{ }}\right)
^{\varepsilon _{n}}(1-e^{-\delta r})_{2}F_{1}(-n,n+2\left( \varepsilon
_{n}+1\right) ;2\varepsilon _{n}+1;e^{-\delta r}),
\end{equation}%
with $\varepsilon _{n}=\frac{\mu Ze^{2}}{\hbar ^{2}\delta \left( n+1\right) }%
-\frac{n+1}{2}$ and $D_{n}$ is another normalization factor$.$ This result
is consistent with the NU method [2]. Further, if we take the shift
parameter $d_{0}=0$ in the present approximation, Eq. (42) reduces to the
form%
\begin{equation}
\psi _{nlm}(\mathbf{r,}\theta ,\phi )=D_{nl}r^{-1}\left( e^{-\delta r\text{ }%
}\right) ^{\varepsilon _{nl}}(1-e^{-\delta r})_{2}^{l+1}F_{1}(-n,n+2\left(
\varepsilon _{nl}+l+1\right) ;2\varepsilon _{nl}+1;e^{-\delta
r})Y_{lm}(\theta ,\phi ),
\end{equation}%
with $\varepsilon _{nl}=\frac{\mu Ze^{2}}{\hbar ^{2}\delta \left(
n+l+1\right) }-\frac{n+l+1}{2}$ and $D_{nl}$ is a normalization factor$.$
The critical screening $\delta _{c}=\frac{2\mu Ze^{2}}{\hbar ^{2}\left(
n+l+1\right) ^{2}}$ at which $E_{nl}=0$ has wave functions: $\psi _{nlm}(%
\mathbf{r,}\theta ,\phi )=D_{nl}r^{-1}(1-e^{-\delta
_{c}r})^{l+1}P_{n}^{(0,2l+1)}(1-2e^{-\delta _{c}r})Y_{lm}(\theta ,\phi ).$

In order to show the accuracy of our analytical results, we present the
numerical data in support of the results obtained in Eqs. (34) and (42)
which are the main analytic results obtained in this work. Therefore, we
calculate the energy eigenvalues for $Z=1,$ $n$ and $l$ arbitrary quantum
numbers as a function of the screening parameters $\delta .$ The results
calculated in Tables 1 and 2 by using Eq. (34) are compared with those
obtained with the help of the numerical integration [37], asymptotic
iteration [5], variational [37], SUSY [3] and the recently proposed
approximation [43] methods. Tables 1 and 2 show that our results obtained
with the new approximation scheme with the NU method are in high agreement
with those obtained by numerical integration method [37] for short potential
range (small screening parameter $\delta $). However, the slight differences
in the energy eigenvalues from the numerical integration method [37] are
observed for long potential range (large screening parameter $\delta $).
Therefore, our approximated numerical results are closer to the numerical
integration results than the results obtained via AIM [5] using Eq. (32) and
also the recently proposed approximation scheme [43] using Eq. (19) for
small and large screening parameter $\delta $ values. Thus, the present
approximation form (22) to the centrifugal term can highly improve the
accuracy of calculating the energy eigenvalues for the Hulth\'{e}n potential
than the recently proposed approximation (5) given in [43].

\section{Normalization of the radial wave function}

Using $s(r)=e^{-\delta r}$ and Eq. (41), we are able to express
normalization condition $\int_{0}^{\infty }u(r)^{2}dr=1$ as%
\begin{equation}
\frac{\mathcal{N}_{nl}^{2}}{\delta }\int_{0}^{1}s^{2\varepsilon
_{nl}-1}(1-s)^{2l+2}\left[ P_{n}^{(2\varepsilon _{nl},2l+1)}(1-2s)\right]
^{2}ds=1.
\end{equation}%
Unfortunately, there is no formula available to calculate this key
integration. Neveretheless, we can find the explicit normalization constant $%
N_{nl}.$ For this purpose, it is not difficult to obtain the results of the
above integral by using the following formulas [56-59],%
\begin{equation}
P_{n}^{(\alpha ,\beta )}(x)=\left( n+\alpha \right) !\left( n+\beta \right)
!\sum\limits_{p=0}^{n}\frac{1}{p!(n+\alpha -p)!\left( \beta +p\right)
!\left( n+p\right) !}\left( \frac{x-1}{2}\right) ^{n-p}\left( \frac{x+1}{2}%
\right) ^{p},
\end{equation}%
and%
\begin{equation}
B(x,y)=\int_{0}^{1}t^{x-1}(1-t)^{y-1}dt=\frac{\Gamma (x)\Gamma (y)}{\Gamma
(x+y)},\text{ }\func{Re}(x),\func{Re}(y)>0.
\end{equation}%
Thus, the normalization constant $\mathcal{N}_{nl}$ is now obtained as%
\begin{equation*}
\mathcal{N}_{nl}=\frac{1}{(n+2l+1)!\Gamma (2\varepsilon _{nl}+n+1)}\sqrt{%
\frac{\delta \Gamma (2\varepsilon _{nl}+2n+2l+4)}{\Gamma (2\varepsilon
_{nl}+2n+1)\sum\limits_{p,q=0}^{n}\left( f_{p}f_{q}f_{p,q}\right) ^{-1}}},
\end{equation*}%
where%
\begin{equation*}
f_{p}=(-1)^{p}p!\Gamma (2\varepsilon _{nl}+n-p+1)(2l+p+1)!\left( n+p\right)
!,
\end{equation*}%
\begin{equation*}
f_{q}=(-1)^{q}q!\Gamma (2\varepsilon _{nl}+n-q+1)\left( 2l+q+1\right)
!\left( n+q\right) !,
\end{equation*}%
\begin{equation}
f_{p,q}=\left( 2l+p+q+2\right) !.
\end{equation}

\section{Conclusions}

In this work, we have proposed an alternative improved approximation scheme
for the centrifugal term and used this approximation scheme together with
the NU method to solve the Schr\"{o}dinger equation with any orbital angular
momentum number $l$ for the Hulth\'{e}n potential. The bound state energy
eigenvalues and the unnormalized radial wavefunctions have been calculated
in analytical and numerical way. The analytic expressions for the energy
eigenvalues and wavefunctions have been reduced to the $s$-wave case and the 
$d_{0}=0$ case (usual approximation) [1-16]. Our numerical results obtained
by the approximation scheme given in expression (22) for the centrifugal
term has been found to be more effective than the numerical results of the
recently proposed approximation (5) of Ref. [43] and the commonly used
approximation in generating the energy spectrum of the Hulth\'{e}n
potential. Our results in Tables 1 and 2 for small screening $\delta $
values show that the present approximation is in high agreement with the
numerical integration and variational methods [37] whereas it is in quite
good agreement for large screening $\delta $ values. The present
approximation method is simple, practical and powerful than the other known
methods [2,5,43]. This new method can be used for many quantum models to
improve the accuracy of energy eigenvalues for few potential models of the
exponential-type like the hyperbolical and Manning-Rosen potentials (cf.
e.g., Refs. [60,61].)

\acknowledgments The author thanks the kind referees for their useful
suggestions. This work was partially supported by the Scientific and
Technological Research Council (T\"{U}B\.{I}TAK) of Turkey.

\newpage

{\normalsize 
}

\bigskip \baselineskip= 2\baselineskip
\bigskip \newpage

\bigskip {\normalsize 
}

\begin{figure}[tbp]
\caption{A plot of the variation of the centrifugal term, $1/r^{2}$ and its
corresponding approximation expression $\protect\delta ^{2}\left[ d_{0}+%
\frac{e^{\protect\delta r}}{\left( e^{\protect\delta r}-1\right) ^{2}}\right]
$\ versus $\protect\delta r,$ where the screening parameter $\protect\delta $
changes from $\protect\delta =0.050$ to $\protect\delta =0.250$ in steps of $%
0.050.$ The parameters are in atomic units $(\hbar =\protect\mu =e=1)$ and $%
Z=1.$ }
\QTP{Dialog Text}
\FRAME{itbpF}{2.527in}{1.9441in}{0in}{}{}{Figure}{\special{language
"Scientific Word";type "GRAPHIC";maintain-aspect-ratio TRUE;display
"USEDEF";valid_file "T";width 2.527in;height 1.9441in;depth
0in;original-width 7.555in;original-height 5.7951in;cropleft "0";croptop
"1";cropright "1";cropbottom "0";tempfilename
'D:/swp40/Docs/KCZT1M00.wmf';tempfile-properties "XTR";}}
\label{1}
\end{figure}

\baselineskip= 2\baselineskip
\bigskip \newpage\ 
\begin{table}[tbp]
\caption{Bound energy spectra of the Hulth\'{e}n potential as a function of
the screening parameter $\protect\delta $ for $2p,$ $3p$ and $3d$ states for 
$Z=1$ in atomic units $(\hbar ^{2}=\protect\mu =e=1)$}%
\begin{tabular}{llllllll}
State & $\delta $ & Present\tablenotemark[1]\tablenotetext[1]{The present
approximation.} & Previous [43]\tablenotemark[2]\tablenotetext[2]{Jia et al
approximation.} & Numerical [37] & AIM [5] & Variational [37] & SUSY [3] \\ 
\tableline$2p$ & $0.025$ & $0.1127611$ & $0.1126344$ & $0.1127605$ & $%
0.1128125$ & $0.1127605$ & $0.1127605$ \\ 
& $0.050$ & $0.1010442$ & $0.1009128$ & $0.1010425$ & $0.1012500$ & $%
0.1010425$ & $0.1010425$ \\ 
& $0.075$ & $0.0898495$ & $0.0898350$ & $0.0898478$ & $0.0903125$ & $%
0.0898478$ & $0.0898478$ \\ 
& $0.100$ & $0.0791769$ & $0.0794011$ & $0.0791794$ & $0.0800000$ & $%
0.0791794$ & $0.0791794$ \\ 
& $0.150$ & $0.0593981$ & $0.0604650$ & $0.0594415$ & $0.0612500$ & $%
0.0594415$ & $0.0594415$ \\ 
& $0.200$ & $0.0417078$ & $0.0441045$ & $0.0418860$ & $0.0450000$ & $%
0.0418860$ & $0.0418854$ \\ 
& $0.250$ & $0.0261059$ & $0.0303195$ & $0.0266111$ & $0.0312500$ & $%
0.0266108$ & $0.0266060$ \\ 
& $0.300$ & $0.0125925$ & $0.0191101$ & $0.0137900$ & $0.0200000$ & $%
0.0137878$ & $0.0137596$ \\ 
& $0.350$ & $0.0011675$ & $0.0104763$ & $0.0037931$ & $0.0112500$ & $%
0.0037734$ & $0.0036146$ \\ 
$3p$ & $0.025$ & $0.0437072$ & $0.0436848$ & $0.0437069$ & $0.0437590$ & $%
0.0437069$ & $0.0437068$ \\ 
& $0.050$ & $0.0331623$ & $0.0332390$ & $0.0331645$ & $0.0333681$ & $%
0.0331645$ & $0.0331632$ \\ 
& $0.075$ & $0.0239207$ & $0.0242183$ & $0.0239397$ & $0.0243837$ & $%
0.0239397$ & $0.0239331$ \\ 
& $0.100$ & $0.0159825$ & $0.0166227$ & $0.0160537$ & $0.0168056$ & $%
0.0160537$ & $0.0160326$ \\ 
& $0.150$ & $0.0040162$ & $0.0057067$ & $0.0044663$ & $0.0058681$ & $%
0.0044660$ & $0.0043599$ \\ 
$3d$ & $0.025$ & $0.0436044$ & $0.0435371$ & $0.0436030$ & $0.0437587$ & $%
0.0436030$ & $0.0436030$ \\ 
& $0.050$ & $0.0327508$ & $0.0329817$ & $0.0327532$ & $0.0333681$ & $%
0.0327532$ & $0.0327532$ \\ 
& $0.075$ & $0.0229948$ & $0.0238893$ & $0.0230307$ & $0.0243837$ & $%
0.0230307$ & $0.0230306$ \\ 
& $0.100$ & $0.0143364$ & $0.0162600$ & $0.0144842$ & $0.0168055$ & $%
0.0144842$ & $0.0144832$ \\ 
& $0.150$ & $0.0003124$ & $0.0053907$ & $0.0013966$ & $0.0058681$ & $%
0.0013894$ & $0.0132820$%
\end{tabular}%
\end{table}

\bigskip

\begin{table}[tbp]
\caption{Bound energy spectra of the Hulth\'{e}n potential as a function of
the screening parameter $\protect\delta $ for $4p,$ $4d,$ $4f,$ $5p,$ $5d,$ $%
5f,$ $5g,$ $6p,$ $6d,$ $6f$ and $6g$ states for $Z=1$ in atomic units $%
(\hbar ^{2}=\protect\mu =e=1)$}%
\begin{tabular}{llllllll}
State & $\delta $ & Present\tablenotemark[1]\tablenotetext[1]{The present
approximation.} & Previous [43]\tablenotemark[2]\tablenotetext[2]{Jia et al
approximation.} & Numerical [37] & AIM [5] & Variational [37] & SUSY [3] \\ 
\tableline$4p$ & $0.025$ & $0.0199486$ & $0.0199625$ & $0.0199489$ & $%
0.0200000$ & $0.0199489$ & 0.0199480 \\ 
& $0.050$ & $0.0110442$ & $0.0111938$ & $0.0110582$ & $0.0112500$ & $%
0.0110582$ & 0.0110430 \\ 
& $0.075$ & $0.0045370$ & $0.0049439$ & $0.0046219$ & $0.0050000$ & $%
0.0046219$ & 0.0045385 \\ 
& $0.100$ & $0.0004269$ & $0.0012128$ & $0.0007550$ & $0.0012500$ & $%
0.0007532$ & 0.0004434 \\ 
$4d$ & $0.025$ & $0.0198457$ & $0.0198877$ & $0.0198462$ & $0.0200000$ & $%
0.0198462$ & 0.0198460 \\ 
& $0.050$ & $0.0106327$ & $0.0110819$ & $0.0106674$ & $0.0112500$ & $%
0.0106674$ & 0.0106609 \\ 
& $0.075$ & $0.0036111$ & $0.0048327$ & $0.0038345$ & $0.0050000$ & $%
0.0038344$ & 0.0037916 \\ 
$4f$ & $0.025$ & $0.0196914$ & $0.0197756$ & $0.0196911$ & $0.0200000$ & 
0.0196911 & 0.0196911 \\ 
& $0.050$ & $0.0100154$ & $0.0109150$ & $0.0100620$ & $0.0112500$ & 0.0100620
& 0.0100618 \\ 
& $0.075$ & $0.0022222$ & $0.0046682$ & $0.0025563$ & $0.0050000$ & 0.0025557
& 0.0025468 \\ 
$5p$ & $0.025$ & $0.0094017$ & $0.0094325$ & $0.0094036$ & $0.0094531$ &  & 
0.0094011 \\ 
& $0.050$ & $0.0026067$ & $0.0027900$ & $0.0026490$ & $0.0028125$ &  & 
0.0026056 \\ 
$5d$ & $0.025$ & $0.0092988$ & $0.0093914$ & $0.0093037$ & $0.0094531$ &  & 
0.0092977 \\ 
& $0.050$ & $0.0021952$ & $0.0027454$ & $0.0023131$ & $0.0028125$ &  & 
0.0022044 \\ 
$5f$ & $0.025$ & $0.0091445$ & $0.0093898$ & $0.0091521$ & $0.0094531$ &  & 
0.0091507 \\ 
& $0.050$ & $0.0015779$ & $0.0026791$ & $0.0017835$ & $0.0028125$ &  & 
0.0017421 \\ 
$5g$ & $0.025$ & $0.0089387$ & $0.0092480$ & $0.0089465$ & $0.0094531$ &  & 
0.0089465 \\ 
& $0.050$ & $0.0007549$ & $0.0025920$ & $0.0010159$ & $0.0028125$ &  & 
0.0010664 \\ 
$6p$ & $0.025$ & $0.0041500$ & $0.0041899$ & $0.0041548$ & $0.0042014$ &  & 
0.0041493 \\ 
$6d$ & $0.025$ & $0.0040471$ & $0.0041671$ & $0.0040606$ & $0.0042014$ &  & 
0.0040452 \\ 
$6f$ & $0.025$ & $0.0038927$ & $0.0042014$ & $0.0039168$ & $0.0042014$ &  & 
0.0038901 \\ 
$6g$ & $0.025$ & $0.0036870$ & $0.0040876$ & $0.0037201$ & $0.0042014$ &  & 
0.0036943%
\end{tabular}%
\end{table}

\bigskip

\end{document}